\def\ga{{\rm\thinspace gauss}}

\def\Msun{\hbox{$\rm\thinspace M_{\odot}$}}

\def\mdot{\hbox{$\dot{m}$}}

\def\spose#1{\hbox to 0pt{#1\hss}}
\def\approxlt{\mathrel{\spose{\lower 3pt\hbox{$\sim$}}
        \raise 2.0pt\hbox{$<$}}}
\def\approxgt{\mathrel{\spose{\lower 3pt\hbox{$\sim$}}
        \raise 2.0pt\hbox{$>$}}}

\documentstyle[psfig]{mn}
\input{psfig.sty}

\newif\ifAMStwofonts

\title[XTE J1118+480]
        {Magnetic flares and the optical variability of the X-ray transient XTE J1118+480}
\author[A. Merloni et al.]
        {A.~Merloni$^1$, T.~Di~Matteo$^{2}$\thanks{{\em Chandra} Fellow} and A.~C.~Fabian$^1$
\\$1$ Institute of Astronomy, Madingley Road, Cambridge, CB3 0HA.
\\$2$ Harvard-Smithsonian Center for Astrophysics, 60 Garden St., Cambridge, MA 02138, USA.
}

\date{}

\begin{document}

\maketitle

\label{firstpage}

\begin{abstract}
The simultaneous presence of a strong quasi periodic oscillation of
period $\sim$10 seconds in the optical and X-ray lightcurves of the
X-ray transient XTE J1118+480 suggests that a significant fraction of
the optical flux originates from the inner part of the accretion flow,
where most of the X-rays are produced. We present a model of magnetic
flares in an accretion disc corona where thermal cyclo-synchrotron
emission contributes significantly to the optical emission, while the
X-rays are produced by inverse Compton scattering of the soft
photons produced by dissipation in the underlying disc and by the
synchrotron process itself. Given the observational constraints, we
estimate the values for the coronal temperature, optical depth and
magnetic field intensity, as well as the accretion rate for the
source. Within our model we predict a correlation between optical and
hard X-ray variability and an anticorrelation between optical and soft
X-rays. We also expect optical variability on flaring timescales
($\sim$ tens of milliseconds), with a power density spectrum similar
to the one observed in the X-ray band. Finally we use both the
available optical/EUV/X-ray spectral energy distribution and the low
frequency time variability to discuss limits on the inner radius of
the optically thick disc.

\end{abstract}


\section{Introduction}

The newly discovered transient X-ray source XTE J1118+480
\cite{R2000}, has already shown a number of remarkable properties.  It
was discovered as a weak source with the RXTE All-Sky Monitor on March
29, 2000 at high galactic latitude ($b\sim 62^{o}$). Subsequent RXTE
pointed observations revealed an energy spectrum typical of black hole
candidates in their {\it hard} state, showing a prominent power-law
component in its X-ray spectrum with a photon index of about 1.8 up to
at least 30 keV.  The source was also observed on March 26 in hard
X-rays by BATSE, up to 120 keV (Wilson \& McCollough 2000). The 13th
magnitude optical counterpart, discovered by Uemura, Kato \& Yamaoka
(2000), exhibit a spectrum fairly typical of X-ray Novae in outburst
(e.g.~Garcia et al.~2000).

Being at such an high galactic latitude, the source suffers from low
interstellar absorption, and consequently has also been observed in
the extreme ultra violet (EUV) by the {\it Extreme Ultraviolet
Explorer (EUVE)}. The {\it EUVE} spectrum and fluxes derived by Hynes
et al. (2000) however are extremely sensitive to the details of the
absorbing column and are therefore subject to large uncertainties.

A strong QPO feature with a frequency $\nu \sim 0.08$ Hz has been found
in the X-ray power density spectrum (PDS) of the source in the first
month of the outburst (March 29 -- May 4, see Revnitsev, Sunyaev \&
Borozdin 2000).  More surprisingly, the optical lightcurves also show
strong flickering on timescales of a few seconds or faster and a
prominent quasi periodic feature with a frequency in agreement with
that of the X-ray QPO. Subsequent HST, RXTE and ASCA observations in
the optical/UV and X-ray band respectively, have also shown that the
QPO is drifting to higher frequencies systematically in both bands,
from $\nu \sim 0.08$ Hz to $\nu \sim 0.11$ Hz later in the month
(Haswell et al. 2000b; Yamaoka, Ueda \& Dotani 2000; Patterson private
communication).

The strikingly similar variability properties observed in both bands
suggest that at least a fraction of the optical/UV flux from the
object should be produced in the same region of the accretion flow
where the X-rays are produced and hence reflect a different aspect of
the same phenomenon. This implies that the modulated optical/UV flux
cannot be produced by thermal emission from the accretion disc
itself; the optical/UV flux from a stellar mass black hole arises from
the outer region of the disc (radii greater than hundreds of 
Schwarzchild radii) where the
X-ray emission is negligible.

Here we show that the presence of such optical/UV variability in the
observations of XTE J1118+480 argues for the presence of significant
thermal synchrotron radiation from a magnetically dominated corona (or
possibly from an advection dominated accretion flow, ADAF, occupying
the inner parts of the flow).

The only other object ever observed to show similar short timescale
variability, in both the optical and  X-ray bands, is the black hole
candidate GX 339-4 in its hard state (Motch et al 1982). For the case
of GX 339-4, typically brighter than XTE J1118+480 in X-rays, it was
also suggested that thermal synchrotron emission should provide a
significant contribution to the optical band (Fabian et al. 1982; Di
Matteo, Celotti \& Fabian 1999).

\section{Cyclo-syncrotron emission from a magnetic corona}

\subsection{The Magnetic field in the corona}

It is now well established that magnetic stresses in accretion discs
are likely to be responsible for the transfer of angular momentum (see
Balbus \& Hawley 1998 for a review). Magneto-rotational instabilities
can drive turbulence by amplifying the seed magnetic fields on roughly
Keplerian timescales; the rate at which magnetic energy is built up is
fast enough to explain the bulk of energy release in an accretion disk
as magnetic dissipation. This supports the idea that accretion disc
coronae, the loci where the hard X-ray emission is produced, are
highly magnetic and form by buoyancy of the strong magnetic fields
amplified in the disk (e.g. Galeev, Rosner \& Vaiana 1979).  This
picture is supported by numerical simulations (Miller \& Stone 2000)
which show that, when weak $B$ fields are amplified via MHD turbulence
in the disk, only a fraction $\sim H/R < 1$ of the energy is
dissipated locally while the rest escapes and forms a strongly
magnetized corona above the disc.

In any such models, the magnetic field in the corona is likely to be
strongly inhomogeneous and to dissipate energy in localized active
regions.  Following Di Matteo, Celotti \& Fabian (1997; 1999), we
assume that a significant fraction, $f$, of the accretion power,
$L=\dot m L_{\rm Edd}$, is accumulated in the corona while the
remaining fraction $(1-f)$ is dissipated internally in the disc. The
magnetic field strength, $B$, in the magnetic flux tubes rising from
the disk can be higher than the values implied by equipartition with
radiation energy density. In fact, if the active region is powered by
the release of magnetic energy, its dissipation velocity is $\sim
v_{\rm A}$, the Alfven speed (see e.g. Di Matteo 1998). The energy
stored in the field is therefore a fraction $v_{A}/c \sim 0.1$ of that
of the radiation \cite{DBF97}.  Based on similar arguments Haardt,
Maraschi \& Ghisellini (1994) showed that the magnetic energy is
released on typical timesclales $ t_{0} \sim 10 R_{\rm b}/c$, where
$R_{\rm b}$ is the size of the active region. This leads to a magnetic
field strength,
\begin{equation} 
{B^2\over 8\pi}\frac{v_{\rm A}}{c}\approx
 {9fL \over 16\pi N (r_{\rm b} R_{\rm s})^2 c} \,
\end{equation}
where $N$ is the number of active regions and $r_{\rm b}$ their dimension in units of  $R_{\rm s}=2 GM / c^2$,
the Schwarzschild radius for a given central mass $M=m \Msun$.
This implies a value
\begin{equation}
B\approx 10^8 {1 \over r_{\rm b}} \left( \frac{\dot m fc}{N m v_{\rm
A}}\right)^{1/2} {\rm G}
\end{equation}
for the estimated magnetic field in the inner part of the accretion flow.

\subsection{Synchrotron emission}

In the hot ($T \sim 50 - 200$ keV), highly magnetized, coronal plasma the
majority of the electrons are thought to be thermal and are expected
to produce synchrotron emission, and Compton scatter the softer photons
produced both by dissipation in the underlying accretion disk and by
the synchrotron processes themselves.

In a thermal plasma, optically--thin synchrotron emission rises
steeply with decreasing frequency. Under most circumstances the
emission becomes self-absorbed and gives rise to a black-body spectrum
below a critical frequency: $\nu_{\rm c} \approx 2.7 \times 10^{14}
\tau^{0.05}(\frac{T}{10^{9} \rm {K}})^{0.95} (\frac{B}{10^6 \rm
{G}})^{0.91}$ Hz (see Wardzi\'nski \& Zdziarski 2000), where $T$ is
the coronal temperature and $\tau$ the scattering optical depth.
Above this frequency, which is right in the optical/UV band for
typical parameter of X-ray transient sources and magnetic fields
$B\sim 10^6-10^7$ G, the cyclo-synchrotron emission decays
exponentially as expected from a thermal plasma, due to the
superposition of cyclotron harmonics.

In an active region above an accretion disc, the relative importance
of the cyclo-synchrotron emission (and its Comptonization) as opposed
to the Comptonization of the external photon field has to be estimated
by comparing the relevant timescales for the processes.  Such
comparisons have shown that, as the synchrotron emission is heavily
self-absorbed, the internal energy density of the synchrotron
radiation $U_{\rm rad,\,int} < B^2/8\pi \approxlt U_{\rm rad,\,ext}$, where
$U_{\rm rad, ext}=(1-f)L/4\pi c R_{\rm disc}^2$ is the `external'
energy density of the radiation emerging from the disc. Consequently
(and in accordance to standard models) most of the X-ray emission in
the soft and hard states of GBH is usually due to Comptonization of
the external photon fields (and not of the
synchrotron component; Di Matteo, Celotti \& Fabian 1997; 1999;
Wardzi\'nski \& Zdziarski 2000).  Nevertheless, in the hard states of
GBH the disk blackbody component, which dominates the X-ray emission
in the soft state, often disappears or cools down significantly, and
the synchrotron component is likely to contribute at least in the
optical/UV band.

Here we show that thermal synchrotron emission can be an important
radiation process in the case of XTE J1118+480.


\section{Application to XTE J1118+480}
\subsection{Spectral modeling}
Hubble Space Telescope observations of XTE J1118+480 (Haswell et
al. 2000a) were made on 2000 April 8, at the end of the rising phase
of the outburst. The data indicate a flux density at $1500 \rm {\AA}$
of $3 \times 10^{-13} {\rm ergs}\; {\rm cm}^{-2} {\rm s}^{-1} {\rm
\AA}^{-1}$ and a spectrum between 1200 and 8000 $\rm {\AA}$ slightly
flatter than $F_{\nu} \propto \nu^{1/3}$.  On the other hand,
unfiltered CCD fast photometry \cite{H2000b,P2000} has revealed
optical flickering of $\pm 0.2$ mag on timescales of seconds, which
correspond to $\sim 20$ per cent of the optical flux.  As the optical
variability is strongly modulated by a QPO with a frequency
comparable (if not the same) to that of the X-ray one, it
has to be produced where most of the X-rays are produced, namely in
the inner part of the accretion flow.

We assume that this rapid variability is due to self absorbed
cyclo-synchrotron (CS) emission. To explicitly show the various
parameter dependences of the synchrotron emission we write the
flux at the peak frequency $\nu_{\rm c}$ as \cite{DCF97}
\begin{eqnarray}
\label{fcs}
F_{\rm CS}&=&2\pi m_{e} \nu_c^3 \theta r_{\rm b}^2 N \left(\frac{R_{\rm S}}{D}\right)^2 \left[\frac{1}{1+t_{\rm CS}/t_{\rm iC}}\right] \\ \nonumber
&\simeq&5.7 \times 10^{-14} \theta  r_{\rm b}^2 N \left(\frac{\nu_c}{10^{15}}\right)\left(\frac{m}{d}\right)^2 \\ \nonumber
&\times& \left[\frac{1}{1+3.2 \times 10^2 \tau U_{\rm rad} \left(\frac{10^{15}}{\nu_c}\right)^3}\right] , \nonumber
\end{eqnarray}
where $d$ is the distance of the source in kiloparsecs,
$\theta=kT/m_ec^2$ the dimensionless temperature, $U_{\rm rad} =
U_{rad, int} + U_{\rm rad, ext}$ and the ratio of the relevant
timescales for the cyclo-synchrotron and inverse Compton emission
determines which process dominates in an active (see Section 2.2).
Under the assumption that at least one fourth of the flux at
$\nu=\nu_c=10^{15}$ Hz is due to CS, we deduce that the intrinsic
dissipation in the accretion disc has to be fairly small ($\dot m
(1-f)\sim 10^{-3}$) implying a magnetic field intensity $B \simeq 2
\times 10^6$ G.


We model the spectrum of the source following the work of Di Matteo,
Celotti \& Fabian (1999). The main feature of the model is that it
considers reprocessing of coronal radiation in the accretion disc not
only according to the active blob size but also according to their
height above the disc. The temperature and the optical depth in the
corona are constrained from the observed slope of the power-law in the
X-ray spectrum ($\alpha\simeq0.8 \propto (\tau \theta)^{-1}$; e.g.
Wardzi\'nski \& Zdziarski 2000). The spectrum is calculated
self-consistently for every annulus of radius $R=rR_{\rm S}$ and width
$dR \ll R$ and integrated over radius from $r=3$ to $r=1000$.  We take
into account all the relevant radiative processes and rescale them
opportunely according to the ratios of their typical cooling
timescales which are obtained by integrating each spectral component
(more accurately than in Eq.~3)

In all cases we postulate that the accretion disc extends down to the
innermost stable orbit of a non-rotating black hole ($R_{\rm
in}=3R_{\rm S}$), so that both the optical and X-rays are produced
where most of the energy is dissipated.  The number of active regions
at any given annulus is calculated as in Haardt, Maraschi \&
Ghisellini (1994), following their discussion on the timescales over
which the magnetic field is amplified in the disc ($t_{\rm amp}\sim
t_{\rm Kep}$) and released in the corona ($t_{\rm rel}\sim t_0$, see
section 2.1). We obtain $dN=9.5 r^{-3/2} dr$, so that the total number
of flares active at any time $N \simeq 10$, as usually assumed from
variability arguments. The mass is fixed to $10$ solar masses.

In Figure 1 we show an illustration of two model spectra for the
recent optical and X-ray observations of XTE J1118+480. They are not
to be regarded as detailed fits to the data, rather as indicative of
the physical properties of the source implied by the model. In particular, we
do not attempt to fit the optical continuum with a realistic model for the 
outermost parts of the accretion disc, our main interest being to put constraints on the nature of the inner accretion flow. 
The relatively high optical-to-X-ray flux ratio for the source imposes the
major constraints on the model. In order for both the synchrotron
component to be important in the optical band and the Comptonized flux
not to exceed the observed limit in the hard X-ray band $U_{\rm
rad,\,ext}$, the external radiation energy density intercepting the
active region (see Eq.~3) has to be lower than $ (1-f)L/(4 \pi R_{\rm
disc}^2 c)$.

The first spectrum (case a, solid lines) is obtained
assuming static dissipation regions at fixed height.  For $r_{\rm
b}=3$ we
have to allow the reconnection sites to be above the accretion disc at
a height of $\sim 4-5 r_{\rm b}$. The need for such aspect ratio
of the active region is just an indication that the soft radiation
field seen by the flaring region has been reduced.

As expected from the simple scaling arguments of Eqs.~(2) and (3), we
obtain a relatively low accretion rate ($\mdot\simeq 0.01$) and a high
value of the fraction of the power dissipated in the corona
($f\simeq0.97$), both consistent with the source being in its {\it
hard} state. However, in this geometry $f$ is not a relevant
parameter because the hard X-ray radiation illuminates the disk and a
substantial fraction of it ($\la 0.5$, depending on the disc albedo)
is reprocessed and thermalized giving rise to a hotter blackbody-like
component roughly similar to that obtained for lower values of $f$.

In Figure 1 (case b, dashed lines) we illustrate an alternative way to
reduce the contribution from this thermal component in the soft X-ray
band and enhance the the cyclo-synchrotron emission in the optical/UV
band. Following Beloborodov (1999), we allow for bulk relativistic
motion of the emitting coronal plasma away from the disc. Due to
Doppler boosting the radiation from a reconnection site towards the
disc is reduced. This is equivalent, for our purpose, to the
assumption that the coronal emission is anisotropic: for $v/c \sim
0.3$, roughly $90$ per cent of it is emitted upwards and only the remaining 
$\sim 10$ per cent impinges on the underlying disc.

Finally, the values of temperature and optical depth of the active
regions we derive are, respectively, $\theta \simeq 0.3$ and
$\tau \simeq 0.6$ and are almost equal in the two cases.  The distance
we obtain is of the order of $D=0.4$ kpc, as expected for a high
latitude galactic source.

\begin{figure}
\vbox to150mm{\vfil 
\psfig{figure=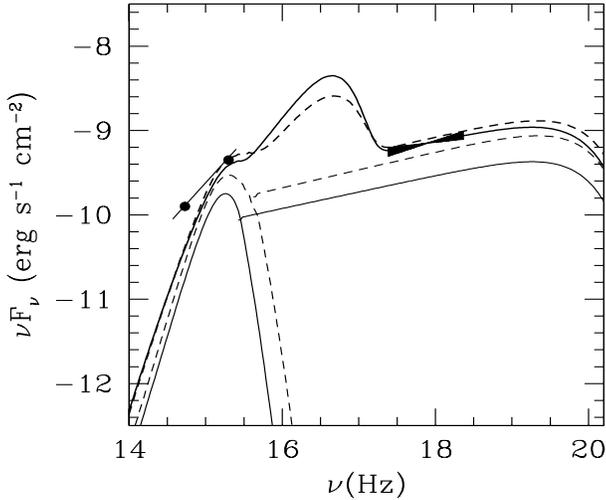,angle=0,height=80mm,width=90mm}
\caption{Representative model predictions for the spectral energy distribution of XTE J1118+480. The points represent optical (V band) (Uemura et al. 2000), 
near UV (Haswell et al. 2000a) and X-ray (Remillard et al. 2000,
Yamaoka et al. 2000) observations of the source during its outburst.
The two thicker lines represent models with (case a, solid line) and
without (case b, dotted lines) bulk relativistic motion of the active
regions.  The parameters are very similar for the two cases: for the
active region parameters we obtain, respectively $\tau=0.61$,
$\theta=0.3$, $r_{\rm b}=2.9$, $h=5.3r_{\rm b}$ (case a) and
$\tau=0.56$, $\theta=0.33$, $r_{\rm b}=3.2$, $h=4r_{\rm b}$ (case b).
We have assumed a mass of $10 \Msun$, and we estimate a distance of
0.4 kpc; the accretion flow parameters are: $\dot m=0.01$ and
$f=0.97$.  The self absorbed cyclo-synchrotron emission and its
Comptonization are plotted (thinner lines).  The optically thick part
significantly contributes to the optical emission, together with the
expected (non-varying) emission from the outermost parts of the
accretion disc, while the X-ray are mainly produced by Comptonization
of disc photons.}
\vfil}
\label{fig1}
\end{figure}

\subsection{The inner disc}

Hynes ey al. (2000) have recently reported on EUV observations of XTE
J1118+480 and have shown that the relatively low EUV flux may be
inconsistent with an optically thick disc extending down to radii
$r_{\rm in}
\approxlt 1000$. However EUV observations are extremely sensitive to
the assumed absorption and the derived limits on $r_{\rm in}$ highly
uncertain. This is further emphasized by recent {\it ASCA}
observations which appear to show a slight soft excess below 2 keV
(Yamaoka et al. 2000) well fitted by a multicolour disc of temperature
0.2 $\pm$ 0.1 keV. Because of the inconsistency between the ASCA and
EUVE measurements we did not use the EUVE data as a further constraint
for our model.

However, in accordance with the requirements discussed above, most
 objects observed in their hard/low state often show either no or very
 little evidence for soft blackbody emission or strong reflection
 features (i.e., the backscattered emission from the putative
 accretion disk). Based on this, it has therefore been argued that, in
 the hard state, geometrically thin discs may not extend down to the
 innermost stable orbit but are truncated at tens or hundreds of
 Schwarzchild radii.  (Gierli\'nsky et al.\ 1997; Poutanen
\& Coppi 1998; Zdziarski et al.\ 1998; Done \& \.Zycki 1998; Esin et
al.\ 1997; Wilms et al.\ 1999 and references therein).

If the disc does not extends down to the innermost stable orbit, then
$U_{\rm rad, ext}$ is suppressed and the relative importance of the
synchrotron component (see Eq.~3) increases, as required by the data. 
Therefore, a central advection
dominated (ADAF) component of the flow extending out to hundreds or
thousands of Schwarzschild radii and surrounded by an optically thick
accretion disc may also fit the spectrum reasonably well (see
e.g. Esin et al. 1998).  The information provided by temporal studies
are then crucial for constraining the geometry of the inner disk, in
particular optical-X-rays cross-correlation and time lag analysis for
the QPO component.

\subsubsection{The $0.1$ Hz QPO}

Psaltis, Belloni \& van der Klis (1999) have suggested that the
frequencies of the periodic features in the PDS of black hole
candidates (which can appear more or less broadened) follow a quite
tight correlation that encompasses also neutron star sources, and therefore
has to be related to the properties of the accretion flow. Recently
Psaltis \& Norman (2000) have proposed a model in which the QPOs are
produced in a narrow ring in the accretion disc where a discontinuity
in some properties of the flow occurs. The frequencies themselves are
essentially determined by the relativistic proper frequencies of the
system (Stella, Vietri \& Morsink 1999; Merloni et
al. 1999). Assuming such a model, the data can be used to constrain
the position of the flow discontinuity (Di Matteo \& Psaltis 1999).

The published PDS of XTE J1118+480 strongly resembles that of GX 339-4
in the {\it hard} state (see Nowak 2000 for a discussion of the
various variability components). Apart from the strongly peaked QPO at
0.08 Hz, it exhibits a broader feature around 1 Hz. Plotted one
against the other, these frequencies fall remarkably well on the
correlation presented by Psaltis, Belloni \& van der Klis (1999).
This would suggest that the low frequency QPO is analogous to the HBO
in neutron stars and is related to twice the nodal precession
general-relativistic frequency of a perturbed orbit around the compact
object. The value of the transition radius, where the modulation is
produced, depends on the spin of the black hole, but is limited
in the range $25 \ga R_t/R_{\rm S} \ga 6$ for a $10\Msun$ black hole
with dimensionless angular momentum $0.99\ga a \ga 0.01$. This
provides an upper bound for the inner extent of the geometrically thin
accretion disc, which is in conflict with the limits derived from EUVE
observations (Hynes et al. 2000) and significantly reduces the
relevance of a central ADAF component, favoring instead models
discussed in Section 3.1. We note, however, that if the observed 
QPO is associated with the HBO
frequency it is not clear why the expected modulation at the keplerian
frequency at the same radius (with frequency $\nu_{\rm Kep}\sim 30$ Hz) 
is not directly observed,
even though it should have a higher quality factor (although, see
e.g. Nowak 2000).

Alternatively, we can ignore the very broad feature in the PDS at
$\sim 1$ Hz and consider the sharp low frequency QPO as due to
Keplerian modulation of the flow far away from the source.  This would
imply a transition radius $R_t\simeq 500 R_{\rm S}$. Although this
interpretation for the observed variability is less plausible, it may
be in better agreement with the implied EUV fluxes and possibly with
the presence of a central ADAF component. However, in this case it
remains to be explained how the modulation propagates inward in the
advective flow without loss of coherence, given that, in the ADAF
model, the bulk of the optical and X-ray are produced in the inner
part of the flow, where the magnetic field is high enough ($B_{ADAF} 
\sim 10^7 \alpha_{0.1}^{-1/2} m_{10}^{-1/2} \mdot^{1/2}_{0.01} 
r_{3}^{-5/4}$ G) to power the cyclo-synchrotron emission, which is then
Componized to give rise to the high energy spectrum (see e.~g. Narayan
\& Yi 1995; Quataert \& Narayan 1999).

\section{Discussion}
 
We have applied a model for the emission from a highly magnetic,
structured corona to the optical and X-ray observations of the newly
discovered X-ray transient XTE J1118+480.

This source is unique for a number of reasons. It is located at high
galactic latitude and has a very high optical-to-X-ray flux ratio. The
source is observed in the typical black hole candidates hard state
(photon index $\Gamma\simeq 1.8$).  Given the optical/UV to X-ray flux
ratio we have derived constraints for the approximate size, optical
depth and magnetic field strengths of the coronal active regions.

The main point of our work is that the simultaneous presence of a
strong quasi periodic feature in the optical and X-ray lightcurves clearly
suggests that the fluxes in the two bands both originate from the same
region in the inner part of the accretion flow. Self-absorbed
cyclo-synchrotron emission is the natural candidate to explain the
optical variability.
Such emission is expected in any magnetic corona model, and the
inferred magnetic field value ($B\approx 2 \times 10^6$ G) is the one
predicted to arise when the source is in the {\it hard} state (low
value of $\dot m$ and high value of $f$).

Our model is not unique. The relatively high optical-to-X-ray flux ratio observed can be explained either by a strong, structured magnetic corona with a relatvely high scaleheight, or by a central ADAF surrounded by an optically thick accretion disc. 
We have shown that even if cyclo-synchrotron emission plays an
important role in the optical band in both cases, 
in our model the X-ray photons are mainly
produced via inverse Compton scattering of the soft disc
photons, while in the ADAF model the CS emission itself acts as source field.
We gave an example of how variability data can be used to further discriminate 
between the models. 

Independently of the dominant source of soft photons for
Comptonization, we expect that any (small) variation in the physical
parameters characterizing the active coronal regions (either in
temperature or optical depths) driven by a flow discontinuity in the
disk should modulate the emission simultaneously in the optical (via
the CS emission) and in the X-ray band (via inverse Compton
emission). In particular, any increase (at a level of 20 per cent or
so, as required) in temperature or density in the corona will cause an
increase in the cyclo-synchrotron component at optical frequencies,
while at the same time it will make the X-ray spectrum harder. From
any such model we then expect a correlation between the optical and
hard X-ray variability but an anticorrelation between optical and soft
X-rays (this has been observed for the case of GX 339-4; see Motch et
al. 1982). Our model also predicts optical variability on flaring
timescales ($\sim$ tens of milliseconds), with a PDS similar to the
one observed in X-rays. Furthermore the variable optical
component is expected to drop out quickly at low frequencies $\nu
< \nu_{\rm c}$ due to the fast decline of the synchrotron emission in
the Rayleigh-Jeans regime. Time-resolved spectroscopy observations 
of such behaviour would support
this model and allow us to place strong constraints on the magnetic
field strength in the corona.

\section{Acknowledgments}
We thank Annalisa Celotti for many useful suggestions and comments on
 the manuscript and Joe Patterson for useful informations on the
 optical variability.  This work was done in the research network
 `Accretion onto black holes, compact stars and protostars', funded by
 the European Commission under contract number ERBFMRX-CT98-0195. AM
 and ACF thank the PPARC and The Royal Society for support,
 respectively. TDM acknowledges support for this work provided by NASA
 through Chandra Fellowship grant number PF8-10005 awarded by the
 Chandra Science Center, which is operated by the Smithsonian
 Astrophysical Observatory for NASA under contract NAS8-39073.

\bsp

\label{lastpage}


\begin{thebibliography}{}
\bibitem[Balbus \& Hawley 1998]{BH98}
Balbus, S. A. \& Hawley, J. F., 1998, Rev. Mod. Phys. 70, 1.

\bibitem[Di Matteo 1998]{DM98}
Di Matteo, T., 1998, MNRAS, 299, L15.

\bibitem[Di Matteo, Blackman \& Fabian 1997]{DBF97}
Di Matteo, T., Blackman, E. G. \& Fabian, A. C., 1997, MNRAS, 291, L23.
 
\bibitem[Di Matteo, Celotti \& Fabian 1997]{DCF97}
Di Matteo, T., Celotti, A. \& Fabian, A. C., 1997, MNRAS, 291, 805.
 
\bibitem[Di Matteo, Celotti \& Fabian 1999]{DCF99}
Di Matteo, T., Celotti, A. \& Fabian, A. C., 1999, MNRAS, 304, 809.

\bibitem[Di Matteo \& Psaltis 1999]{DP99}
Di Matteo, T. \& Psaltis, D., 1999, ApJ, 526, L101.

\bibitem[Done \& Zycki 1999]{DZ99}
Done, C., \& Zycki, P. T., 1999, MNRAS, 305, 457.

\bibitem[Fabian et al. 1982]{F82}
Fabian, A., Guilbert, P. W., Motch, C., Ricketts, M., Ilovaisky, S. A., Chevalier, C., 1982, A\&A, 111, L9.

\bibitem[Galeev, Rosner \& Vaiana 1979]{GRV79}
Galeev, A. A., Rosner, R. \& Vaiana, G. S., 1979, ApJ, 229, 318.

\bibitem[]{}Garcia M., Brown W., Pahre M., J. McClintock 2000, IAUC 7392.

\bibitem[Gierli\'nski et al. 1997]{Gi97}
Gierli\'nski, M., Zdziarski, A. A., Done, C.,
Johnson, W. N., Ebisawa, K., Ueda, Y., Phlips, F., 1997, MNRAS, 288, 958.

\bibitem[Haardt, Maraschi \& Ghisellini]{HMG94}
Haardt, F., Maraschi, L. \& Ghisellini, G., 1994, ApJ, 432, L95.

\bibitem[Haswell et al. 2000a]{H2000a}
Haswell, Hynes, R. I. \& King, A. R., 2000a, IAUC 742

\bibitem[Haswell et al. 2000b]{H2000b}
Haswell, C. A., Skillman, D., Patterson,J., Hynes, R. I., Cui, W., 2000b, IAUC 7427 

\bibitem[Hynes et al. 2000]{Hy2000}
Hynes, R. I., Mauche, C. W., Haswell, C. A., Schrader, C. R., Cui, W., Chaty, S., 2000, submitted to ApJL, astro-ph/0005398. 

\bibitem[Merloni et al. 1999]{me99}
Merloni, A., Stella, L., Vietri, M., Bini, D., 1999, MNRAS, 304, 155. 

\bibitem[Miller \& Stone 2000]{MS2000}
Miller, K. \& A.Stone, J. M., 2000, ApJ, 534, 398.

\bibitem[Motch et al. 1983]{M83}
Motch, C., Ricketts, M., Page, C. G., Ilovaisky, S. A., Chevalier, C., 1983, A\&A, 119, 171.

\bibitem[Narayan \& Yi 1995]{NY95}
Narayan, R. \& Yi, I., 1995, ApJ, 452, 710.

\bibitem[Nowak 2000]{N2000}
Nowak, M. A., 2000, MNRAS, in press, astro-ph/0005232.

\bibitem[Patterson 2000]{P2000}
Patterson, J., 2000, IAUC 7413; see also http://www.astro.bio2.edu/cba.

\bibitem[Psaltis, Belloni \& van der Klis 1999]{PBv99}
Psaltis, D., Belloni, T. \& van der Klis, M., 1999, ApJ, 520, 262.

\bibitem[Psaltis \& Norman 2000]{PN2000}
Psaltis, D. \& Norman, C., 2000, ApJ, in press, astro-ph/0001391.

\bibitem[Poutanen \& Coppi 1998]{PC98}
Poutanen, J., \& Coppi, P., 1998, Phys. Scripta, T77, 57 (astro-ph/9711316).

\bibitem[Quataert \& Narayan 1999]{QN99}
Quataert, E. \& Narayan, R., 1999, ApJ, 520, 298.

\bibitem[Remillard et al. 2000]{R2000}
Remillard R., Morgan E., Smith D, Smith E. 2000, IAUC 7389

\bibitem[Revnivtsev, Trudolyubov \& Borozdin 2000]{we_1748} 
Revnivtsev M., Sunyaev  R. \& Borozdin K., 2000, submitted to A\&AL, 
astro-ph/0005212. 

\bibitem[Stella, Vietri \& Morsink 1999]{SVM99}
Stella, L., Vietri, M. \& Morsink, S., 1999, ApJ, 524, L63.

\bibitem[Uemura, Kato \& Yamaoka 2000]{UKY2000}
Uemura M., Kato T. \& Yamaoka H. 2000, IAUC 7390.

\bibitem[]{}Wilson C. \& McCollough M. 2000, IAUC 7390

\bibitem[Yamaoka et al. 2000]{Y2000}
Yamaoka, K, Ueda, Y. \& Dotani, T., 2000, IAUC 7427.

\bibitem[Wardzi\'nski \& Zdziarski 2000]{WZ2000}
Wardzi\'nski, G. \& Zdziarski, A. A., 2000, MNRAS, 314, 183.

\bibitem[Wilms et al. 1999]{W99}
Wilms, J., Nowak, M. A., Dove, J. B., Fender,R. P., 
\& Di Matteo, T.,  1999, ApJ, 522, 460.

\bibitem[Zdziarski et al. 1998]{Z98}
Zdziarski, A. A., Poutanen, J., Mikolajewska, J.,
Gierli\'nski, M., Ebisawa, K., \& Johnson, W. N., 1998, MNRAS, 301, 435.


\end{thebibliography}
\end{document}